\documentclass{article} 
\usepackage{iclr2025_conference,times}

\usepackage{amsmath,amsfonts,bm}

\def\eqref#1{equation~\ref{#1}}

\def\1{\bm{1}}

\DeclareMathAlphabet{\mathsfit}{\encodingdefault}{\sfdefault}{m}{sl}
\SetMathAlphabet{\mathsfit}{bold}{\encodingdefault}{\sfdefault}{bx}{n}

\usepackage{hyperref}
\usepackage{graphicx} 
\usepackage{url}

\title{Helix-mRNA: A Hybrid Foundation Model For Full Sequence mRNA Therapeutics}

\author{Matthew Wood \\
Helical\\
Luxembourg \\
\texttt{\small matthew@helical-ai.com} \\
\And
Mathieu Klop \\
Helical \\
Luxembourg \\
\texttt{\small mathieu@helical-ai.com}
\And
Maxime Allard
\thanks{Corresponding Author} \\
Helical \\
Luxembourg \\
\texttt{\small maxime@helical-ai.com}
}

\iclrfinalcopy 
\begin{document}

\maketitle

\begin{abstract}
mRNA-based vaccines have become a major focus in the pharmaceutical industry. The coding sequence as well as the Untranslated Regions (UTRs) of an mRNA can strongly influence translation efficiency, stability, degradation, and other factors that collectively determine a vaccine’s effectiveness. However, optimizing mRNA sequences for those properties remains a complex challenge. Existing deep learning models often focus solely on coding region optimization, overlooking the UTRs. We present Helix-mRNA, a structured state-space-based and attention hybrid model to address these challenges. In addition to a first pre-training, a second pre-training stage allows us to specialise the model with high-quality data. We employ single nucleotide tokenization of mRNA sequences with codon separation, ensuring prior biological and structural information from the original mRNA sequence is not lost. Our model, Helix-mRNA, outperforms existing methods in analysing both UTRs and coding region properties. It can process sequences 6x longer than current approaches while using only 10\% of the parameters of existing foundation models. Its predictive capabilities extend to all mRNA regions.
We open-source the model (\url{https://github.com/helicalAI/helical}) and model weights (\url{https://huggingface.co/helical-ai/helix-mRNA}).
\end{abstract}
\section {Introduction}
Engineered messenger Ribonucleic Acids (mRNAs) have emerged as a versatile platform for advanced therapeutics and vaccines, offering rapid design, scalable production, and flexible administration  \cite{pardi2018mrna}. The swift deployment of COVID-19 mRNA vaccines exemplified these advantages, reducing conventional vaccine development timelines from years to months \cite{park2021covid}. Beyond infectious disease, mRNA-based immunotherapies have shown durable anti-tumour efficacy in cancer \cite{sahin2017personalized}. Further, optimizing mRNA for stability and translation efficiency enables high-yield protein expression in both microbial and mammalian hosts \cite{eisenhut2020systematic}. These advances underscore the broad applicability of mRNA engineering for biomedical and industrial needs, ranging from gene therapies to large-scale protein manufacturing. Deep learning methods have shown promise in further accelerating mRNA therapeutic and vaccine design \cite{castillo2024optimus}, spanning traditional architectures to foundation models, which are models trained on large datasets that serve as building blocks for different downstream tasks\cite{yazdanijahromi2024helm}. 

Current approaches face three key limitations that particularly impact mRNA sequence analysis. First, the short context length used by current approaches is problematic because mRNA sequences vary dramatically in length, from only a few nucleotides to several thousand, making it difficult to capture long-range dependencies effectively. Second, simplified tokenization schemes risk losing important biological structures that exist in the original sequences \cite{yazdanijahromi2024helm}. Third, existing models like HELM \cite{yazdanijahromi2024helm}, CodonBERT \cite{codonbert2024}, and Optimus 5-Prime \cite{castillo2024optimus} are specialised for specific mRNA regions, making them inflexible and inefficient to adapt to other mRNA regions without complete retraining. 

To address these limitations, we introduce Helix-mRNA, a hybrid (attention-based~\cite{vaswani2023attentionneed} and state-space-based~\cite{gu2022s4,gu2024mambalineartimesequencemodeling}) foundation model capable of capturing full-length mRNA sequences. We employ a two-stage pre-training method that addresses task-specific specialisation challenges. Our approach features single nucleotide tokenization with a structural codon representation. Helix-mRNA outperforms current models across various downstream benchmarks. These tasks include prediction of translation efficiency, stability and degradation from UTR and codon information. 

\section{Methods}
\label{method}
We break down the different components of Helix-mRNA and how each of them overcome current challenges in the mRNA modality. Helix-mRNA enables long sequence processing through a hybrid state-based and attention architecture, achieving single nucleotide resolution and still maintaining precise coding region representation. Further, we use a two stage pre-training approach through Warmup-Stable-Decay (WSD) scheduling \cite{hu2024wsd_scheduler}, allowing for a balance between generalisation and specialisation. 

\subsection{Hybrid Architecture}
\label{architecture}

\begin{figure}[h]
\begin{center}
\includegraphics[scale=0.5]{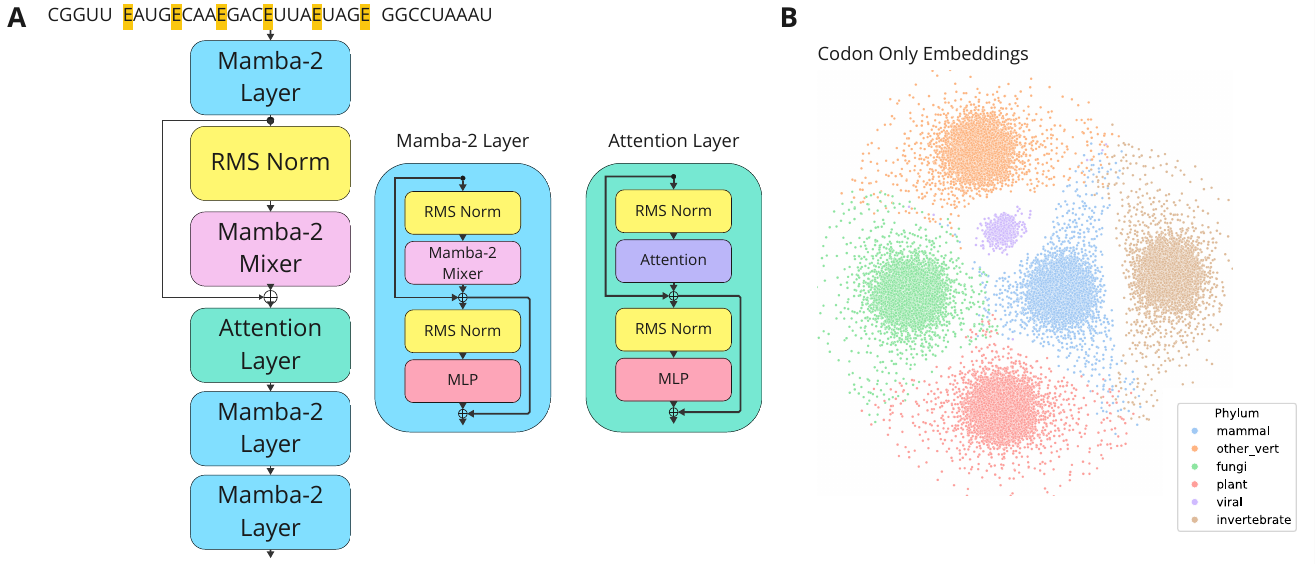}
\end{center}
\caption{\textbf{A)} Helix-mRNA hybrid architecture incorporating state-based and attention-based approaches. We show how we retain coding region structure with an additional token highlighted in the figure. \textbf{B)} Embeddings from initial pre-training, generated using only coding regions from mRNA sequences not seen during training.}
\label{architecture_emb_multi_figure}
\end{figure}

We train a 5.19 million parameter model, 10\% of the parameters of Transformer HELM \cite{yazdanijahromi2024helm}, with a hybrid SSM and attention-based architecture shown in Figure \ref{architecture_emb_multi_figure} taking inspiration from Nemotron \cite{parmar2024nemotron415btechnicalreport}, Jamba \cite{lieber2024jambahybridtransformermambalanguage}, Zamba \cite{glorioso2024zambacompact7bssm} and Samba \cite{ren2024samba}. The ratio of Mamba-2 layers, MLP layers, and attention layers was selected from the different ablation studies performed by \cite{nvidia2024hybrid} with 44.4\% Mamba-2 layers, 44.4\% MLP layers and 11.1\% attention layers. By preceding attention layers with Mamba-2 layers~, the recurrent nature of the Mamba-2 layer encodes positional information needed by the attention layer \cite{vaswani2023attentionneed,dao2024mamba2}. This architectural approach enables Helix-mRNA to read long mRNA sequences with high granularity, combining state-based long sequence handling with the in-context retention capabilities of attention-based methods.

\subsection{Single Nucleotide and Codon Encoding }
\label{tokenization}
Long mRNA sequences are traditionally computationally challenging, addressed by analysing subregions of the sequence \cite{yazdanijahromi2024helm} and coarser tokenization methods that group nucleotides together \cite{codonbert2024}. With our hybrid architecture enabling long sequence lengths, we tokenize full-length sequences at single nucleotide resolution. We map each base (\textit{A, C, U,} and \textit{G}) to a unique integer, ensuring the preservation of complete biological information without any data loss. A key focus of our approach is the mRNA coding region, which plays a critical role in translation efficiency, protein expression, and mRNA stability \cite{boel2016codon}, \cite{hanson2018codon}. To emphasize the importance of codon structure and enhance pattern extraction, we introduce a special character \textit{E} to clearly denote codon separation. We pre-train on a maximum input length of 12288 tokens, 6x the context length of Transformer HELM \cite{yazdanijahromi2024helm}. 

\subsection{Diverse mRNA Sequences Across Multiple Phyla}
\label{data_selection}
We use a taxonomically diverse pre-training dataset which was specifically curated to capture both deep evolutionary conservation patterns in eukaryotic sequences and the distinct nucleotide compositions and structural characteristics of viral genomes. The broad phylogenetic coverage enables the model to learn robust representations of sequence features across different evolutionary distances and genomic architectures. We gathered all RefSeq mRNA sequences\footnote{\url{{https://ftp.ncbi.nih.gov/refseq/release/}}} from a diverse set of eukaryotes, including vertebrates (mammals and non-mammals), plants, invertebrates, fungi, and 238 clinically relevant viruses \cite{o2016reference} to pre-train our Helix-mRNA model (See Appendix~\ref{supp:data}).

\subsection{Two-Stage Specialised Pre-Training}
\label{pre-training}
We utilise a two stage autoregressive unsupervised pre-training strategy, guided by the WSD learning rate scheduler \cite{hu2024wsd_scheduler}. In the first stage, the WSD scheduler manages a warm-up phase followed by a stable learning rate, allowing the model to process data of varying quality, creating a general and robust base model. In the second stage, the WSD scheduler transitions into its decay phase, where training focuses exclusively on high-quality data. During the second phase, human only mRNA sequences are used to refine the model for human specific tasks. This two stage process enables efficient learning from mixed-quality data in the initial stage and achieving specialised distillation through the targeted second stage.

\section{Results}\label{results}
\begin{table}[!h]
\label{benchmark_results}
\caption{Benchmark comparisons showing Helix-mRNA, CodonBERT, Transformer HELM and Transformer XE.
Spearman rank correlations are shown for all tasks.}
\begin{center}
\begin{tabular}{lcccc}
\label{benchmark_results}
\textbf{Task} & \textbf{CodonBERT} & \textbf{HELM} & \textbf{XE} & \textbf{Helix-mRNA} 
\\ \hline \\
\text{MLOS Flu Vaccines} & 0.54 & 0.70 & 0.65 & \textbf{0.79$\pm$}0.121 \\
\text{mRFP Expression} & 0.77 & 0.85 & 0.82 & \textbf{0.86$\pm$}0.008 \\
\text{mRNA Stability/iCodon} & 0.35 & \textbf{0.53} & 0.50 & 0.52$\pm$0.004 \\
\text{Tc-Riboswitch} & 0.50 & 0.63 & 0.57 & \textbf{0.64}$\pm$0.033 \\
\text{Vaccine Degradation} & 0.78 & 0.83 & 0.80 & \textbf{0.84$\pm$}0.032 \\
\hline
\end{tabular}
\end{center}
\end{table}

After the first stage of pre-training, we use Uniform Manifold Approximation (UMAP) to visualise the embeddings of unseen coding sequences from each of the different phyla used during pre-training. 
Figure \ref{architecture_emb_multi_figure} shows that Helix-mRNA clearly clusters coding sequences from different phyla through unsupervised pre-training alone. We demonstrate that despite pre-training on full sequences, Helix-mRNA can extract relevant information from different subregions without the need for further training. Viruses cluster very clearly and separately from all other eukaryotic phyla in the embedding space. This distinct separation implies that the model has learned biologically meaningful features, such as differences in codon usage and sequence structure, enabling it to discriminate viral from non-viral sequences.


\subsection{Biological Downstream Tasks}
We first evaluate Helix-mRNA on coding region related property prediction tasks including stability, degradation and translation efficiency. Our results demonstrate that Helix-mRNA outperforms existing approaches like CodonBert, Transformer HELM, and Transformer XE across multiple benchmarks, including an E. coli related tasks even though Helix-mRNA was not pre-trained on any prokaryotic data, further highlighting generalisability to various tasks (See Appendix~\ref{ref:supp_eval}. We assessed performance using the following benchmarks where we use a 70\% training split, a 15\% validation split and a 15\% test split:
\begin{itemize}
    \item mRNA Stability/iCodon : Stability profiles from vertebrates \cite{diez2022icodon}.
    \item mRFP Expression : Protein production levels in E. coli variants \cite{nieuwkoop2023revealing}.
    \item MLOS dataset: Haemagglutinin antigen encodings for flu vaccines \cite{codonbert2024}.
    \item Tc-Riboswitches : Tetracycline riboswitch dimers with switching factor measurements \cite{groher2018tuning}.
    \item SARS-CoV-2 Vaccine Degradation : mRNA sequences optimized for structural features \cite{leppek2022combinatorial}.
\end{itemize}


\subsection{UTR5 Mean Ribosome Load Prediction}
\label{utr5}

\begin{figure}[h]
\begin{center}
\label{optimus_helix}
\includegraphics[scale=0.4]{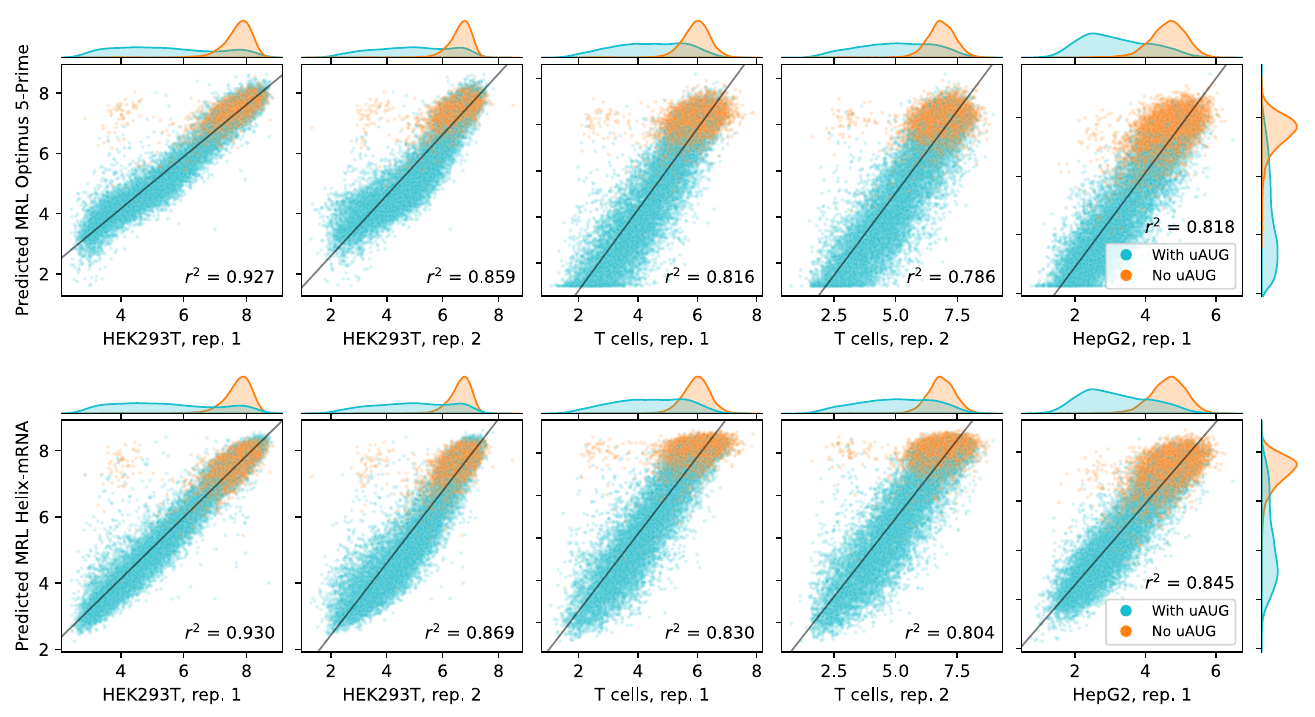}
\end{center}
\caption{Benchmark comparison between Helix-mRNA and Optimus 5-Prime on task specific fine-tuning to predict Mean Ribosome Load (MRL) across 3 cell lines (HEK293T, T cells, and HepG2) using two replicates, reproduced from the Optimus 5-Prime codebase released with the paper \cite{castillo2024optimus}. Results show the $r^2$ correlation values between the predicted MRL and the true MRL.}
\end{figure}
The benchmarks presented in Table \ref{benchmark_results} focus solely on the coding region. However, since Helix-mRNA is capable of utilising both UTR regions and coding regions, we evaluate the model on tasks specific to the 5’ UTR. We evaluated our model’s predictive capabilities by replicating the tasks from \cite{castillo2024optimus}, predicting Mean Ribosome Load (MRL) across three cell lines (HEK293T, T cells, and HepG2) using two replicates, shown in Figure \ref{optimus_helix}. Fine-tuning for 5 epochs enabled us to outperform Optimus 5-Prime on MRL prediction tasks, with the greatest improvements seen in T cells and HepG2 cell lines, where Optimus 5-Prime struggles the most.  By applying our model to their established tasks for cell line specific MRL prediction, we demonstrate our approach’s performance and generalisability to different downstream tasks, further highlighting our model’s versatility.

\section{Conclusion}
In this work we presented Helix-mRNA, a hybrid architecture combining attention mechanisms and SSMs, taking inspiration from recent advances in long-context natural language models \cite{nvidia2024hybrid}, allowing for longer context lengths and better in-context retention capabilities. The extra context length combined with single-nucleotide and codon structure tokenization allows the model to capture additional features useful in mRNA sequence design, namely codons, UTR regions, and secondary structures. We show that our tokenization methods doesn't impact performance and allows us to use the model in coding region only tasks, outperforming Transformer HELM \cite{yazdanijahromi2024helm}, Transformer XE and CodonBERT \cite{codonbert2024} or in UTR 5' region specific tasks by outperforming Optimus 5-Prime \cite{castillo2024optimus}. 
By effectively merging state-of-the-art sequence modelling techniques with two-stage pre-training, Helix-mRNA paves the way for more robust, efficient, and broadly applicable mRNA therapeutics in both clinical and industrial domains.

\subsubsection*{Acknowledgments}
Thanks to the Helical team, especially Issac Goh and Martina Oliver Huidobro, for the discussions around the paper. Thanks also to Prof. Antoine Cully for his support and to LuxProvide for providing computational resources for model pre-training.


\bibliographystyle{iclr2025_conference}

\appendix
\section{Appendix}

\subsection{Specific Architecture Implementation Details}

We use a hidden dimension of 256 throughout the model with SiLU activation functions. For the Mamba-2 layers, we employ a head dimension of 32, a single group, and a kernel size of 4 for convolution, with an expansion factor of 2. The attention implementation uses Flash Attention 2 with 32 attention heads and 8 key-value heads. For MLP layers, we maintain an intermediate size of 512, representing an expansion ratio of 2. No biases are used in the attention and MLP layers. The model consists of 9 total layers, with a layer configuration of 4 Mamba-2 layers, 4 MLP layers and a single attention layer in the configuration \textit{M+M*+M+M+}, where M represents a Mamba-2 layer, + represents an MLP layer and * represents attention layers. 

\begin{table}[h] 
    \centering
    \caption{Model parameters used for Helix-mRNA.}
    \resizebox{\textwidth}{!}{
    \begin{tabular}{lcccccccc}
         Params (M) & No. Layers & Model Dim & Attn. Heads & State Dim. & No. Groups & Pos. Emb. & Seq. Len \\
        \hline \\
        5.19 & 8 & 256 & 16 & 128 & 1 & None & 12288 \\
        \\ \hline \\
    \end{tabular}
    }
\end{table}

\begin{table}[h] 
    \centering
    \caption{Pre-training Details.}
    \resizebox{\textwidth}{!}{
    \begin{tabular}{ccccc}
         No. Tokens (B) & Pre-Training Time (Days) & No. Devices & Device Type  \\
        \hline \\
        332 & 2.3 & 32 & NVIDIA A100-SXM4-40GB   \\
        \\ \hline \\
    \end{tabular}
    }
\end{table}

\subsection{Pre-training Datasets}\label{supp:data}

All of the datasets can be found under the NCBI ftp server: \url{https://ftp.ncbi.nih.gov/refseq/release/}. We downloaded all the files ending by `.rna.gbff.gz` and extracted the utr and coding sequences via our pre-processing pipeline. Out of the 56 Million sequences we subsampled 27 million sequences due to budgetary constrains to train our model. 

The viral component encompassed the main human pathogens, including respiratory viruses (SARS-CoV-2, influenza A/B/C, RSV), retroviruses (HIV-1/2, HTLV-1/2), hepatotropic viruses (HBV, HCV), herpesviruses (HSV-1/2, EBV, VZV), and emerging arboviruses (Zika, Dengue 1-4). The dataset consists of 27 million sequences, with class contributions being, 37.6\% other vertebrates (\url{https://ftp.ncbi.nih.gov/refseq/release/vertebrate_other/}), 24.4\% mammals (\url{https://ftp.ncbi.nih.gov/refseq/release/vertebrate_mammalian/}), 22.8\% invertebrates (\url{https://ftp.ncbi.nih.gov/refseq/release/invertebrate/}), 13.7\% fungi (\url{https://ftp.ncbi.nih.gov/refseq/release/fungi/}) and 1.4\% viruses (\url{https://ftp.ncbi.nih.gov/refseq/release/viral/}).

\subsection{Fine-Tuning}
We unfreeze the 2 last layers when fine-tuning on downstream tasks, while freezing the rest of the parameters.

\subsection{Evaluation Details}\label{ref:supp_eval}
Due to the HELM code not yet being publicly available, we referenced their published research findings directly instead of independently reproducing their experimental methodology in Table \ref{benchmark_results} \cite{yazdanijahromi2024helm}. For our analysis, we selected the highest performance values from either the Masked Language Modelling or Causal Language Modelling variants of the Transformer XE and Transformer HELM models. The authors of HELM report the best results achieved for the tasks, with the exception of the MLOS Flu Vaccines task, where the average performance across three random splits is reported due to the absence of predefined splits. To align with their methodology, we adopt a similar approach but also include an error margin to highlight the variability of results across five runs.

\subsection{Full Sequence Embeddings After Pre-Training}

\begin{figure}[h]
\begin{center}
\label{embedding_figures_full_seq}
\includegraphics[scale=0.3]{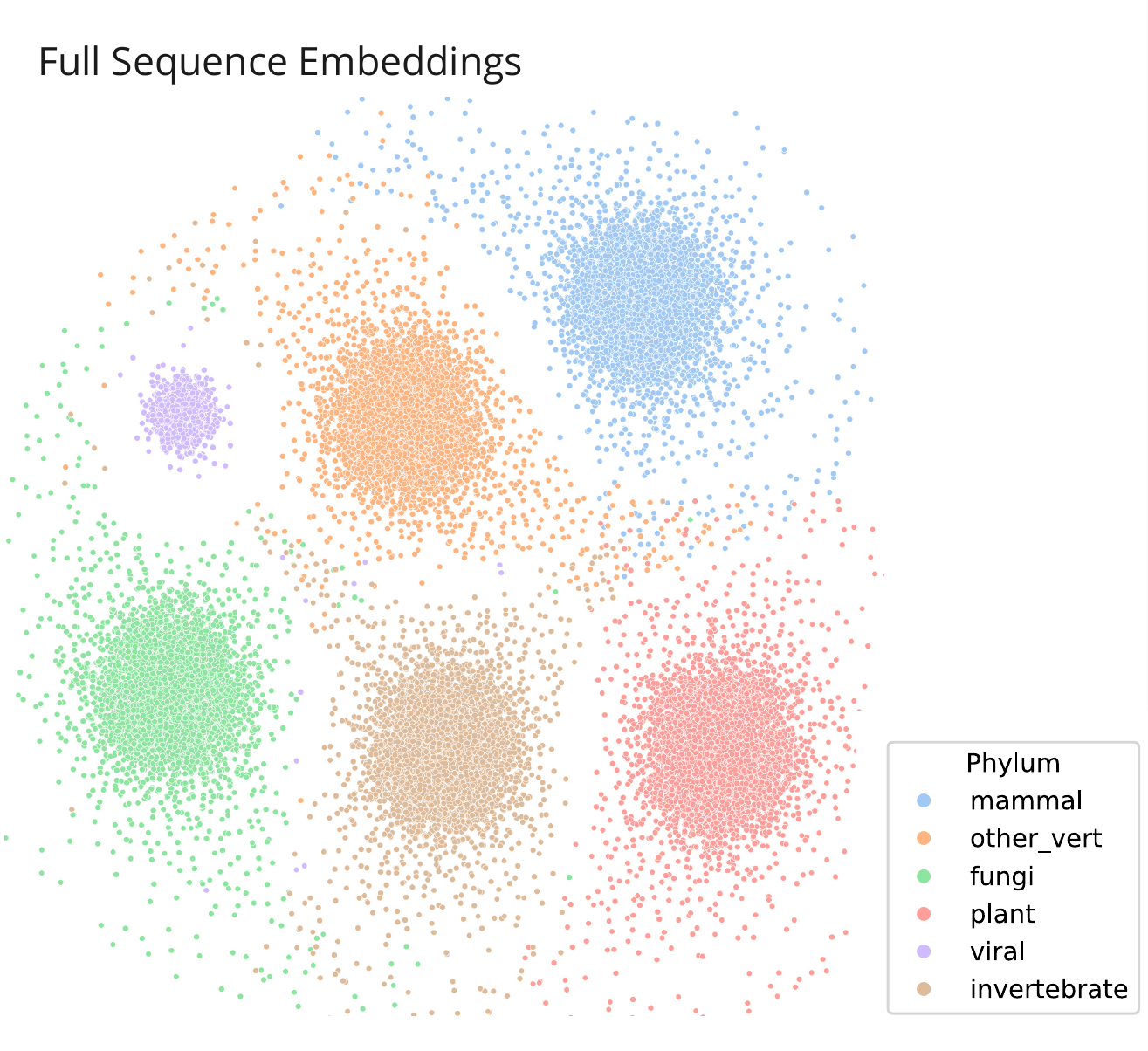}
\end{center}
\caption{Helix-mRNA embeddings from initial pre-training, generated using full mRNA sequences including both the coding and untranslated regions.}
\end{figure}

Embeddings presented in Figure \ref{embedding_figures_full_seq} show Helix-mRNA's ability to process full length sequences not limited to just the coding region.

\end{document}